\documentclass[twocolumn,showpacs,superscriptaddress,prb]{revtex4}
\usepackage{amsfonts}
\usepackage{amsmath}
\usepackage{amssymb}
\usepackage{graphicx}

\begin{document}

\preprint{HEP/123-qed}
\title[Magnetooscillations and negative conductivity]{Microwave-induced
magnetooscillations and absolute negative conductivity in the
multisubband two-dimensional electron system on liquid helium}
\author{Yu.P. Monarkha}
\affiliation{Institute for Low Temperature Physics and Engineering, 47 Lenin Avenue,
61103 Kharkov, Ukraine}

\begin{abstract}
It is shown that a nonequilibrium filling of an upper surface
subband induced by the microwave resonance can be the origin of
the absolute negative conductivity and zero-resistance states for
the two-dimensional electron system on liquid helium under
magnetic field applied normally. Contrary to the similar effect
reported for semiconductor systems, an oscillating sign-changing
correction to the dc-magnetoconductivity appears due to
quasi-elastic iter-subband scattering which does not involve
photons. The analysis given explains remarkable
magnetooscillations and zero-resistance states recently observed
for electrons on liquid helium.
\end{abstract}


\maketitle


Magnetooscillations and absolute negative conductivity induced by microwave
(MW) radiation under magnetic field applied perpendicular to a thin
semiconductor film were predicted theoretically by Ryzhii already in 1969~%
\cite{Ryz-69}. For a long while, these effects were not observed
experimentally. Quite recently, magnetooscillations and the effect of
vanishing resistivity (zero-resistance states) induced by MW radiation were
observed in semiconductor two-dimensional (2D) electron systems of ultrahigh
mobility~\cite{Mani-02}. It was shown that the negative conductivity ($%
\sigma _{xx}<0$) by itself suffices to explain the zero-dc-resistance state
(ZRS) observed~\cite{Andreev-2003}. There are also alternative explanations
which refer to an excitonic mechanism for electron pairing~\cite{Mani-02}.

Recently, microwave-induced magnetooscillations~\cite{KonKon-2009}
and even vanishing resistivity~\cite{KonKon-2009,KonKon-2010} were
observed in the quasi-2D electron system on liquid helium. Despite
striking similarity of these results to those obtained for
semiconductor systems, there are important differences in
experimental conditions. First, it should be emphasized that
surface electrons (SEs) on helium is a nondegenerate system with
strong Coulomb interaction. Therefore, it is doubtful whether
electron paring can be realized in such a system. The second
important difference is that for semiconductor 2D systems
magnetooscillations and ZRS were predicted and observed for a
quite arbitrary MW frequency $\omega $ without excitation of
higher subbands, while for SEs on liquid helium the effect is
observed only when the MW frequency is under the resonance
condition for exciting electrons to the second surface subband:
$\hbar \omega =\Delta _{2}-\Delta _{1}$ (here $ \Delta _{l}$
describes energy spectrum of surface levels, and $l=1,2,...$).
This means that the theory of photon-stimulated impurity
scattering~\cite{Ryz-69} which leads to magnetooscillations and
negative conductivity in semiconductor 2D systems is not
sufficient for explaining ZRS of SEs on liquid helium, at least
under conditions of the experiment.

In this Letter we show that additional electron population of the
first excited subband induced by the MW resonance provides a
possibility for ordinary impurity iter-subband scattering to
result in an oscillating sign-changing correction to $\sigma
_{xx}$. At high enough MW powers and at certain magnetic fields,
this correction leads to absolute negative conductivity and,
according to Ref.~\cite{Andreev-2003}, to the zero-resistance
state observed for SEs on liquid
helium~\cite{KonKon-2009,KonKon-2010}.

In order to understand the physics of appearing of the
sign-changing correction to the dc-magnetoconductivity under the
MW resonance condition, it is instructive to remind the
explanation given for photon-induced impurity
scattering~\cite{Ryz-69}. Consider a multisubband 2D electron
system under perpendicular magnetic field $\mathbf{B}$, and assume
that its energy spectrum is described by $\varepsilon _{l,n}=$
$\Delta _{l}+\hbar \omega _{c}\left( n+1/2\right) $, where $\omega
_{c}$ is the cyclotron frequency $ eB/m_{e}c$ and $n=0,1,2,..$ .
For photon-induced impurity scattering within the ground subband,
electron current $j_{x}$ is proportional to $q_{y}\delta \left[
\hbar \omega _{c}\left( n-n^{\prime }\right) +\hbar \omega
+eEl_{B}^{2}q_{y} \right] $~\cite{Ryz-69}, where $\hbar
\mathbf{q}$ is the momentum exchange, $l_{B}^{2}=\hbar c/eB$, $E$
is the in-plane driving electric field, $\hbar \omega $ is the
photon energy. The delta-function represents the energy
conservation which takes into account Landau level tilting in the
driving field; $eEl_{B}^{2}q_{y}\equiv eE\left( X-X^{\prime
}\right) \equiv q_{y}v_{H} $ (here $v_{H}=cE/B$). Depending on the
sign of $ \hbar \omega _{c}\left( n-n^{\prime }\right) +\hbar
\omega $, the $q_{y}$ could be positive or negative which was the
explanation for changing the direction of the current.

Under the MW resonance condition $\hbar \omega =\Delta _{2}-\Delta _{1}$, a
nonequilibrium filling of the first excited subband gives rise to usual
inter-subband impurity scattering which is restricted by a similar
delta-function $\delta \left[ \hbar \omega _{c}\left( n-n^{\prime }\right)
+\Delta _{2,1}+eEl_{B}^{2}q_{y}\right] $, where $\Delta _{l,l^{\prime
}}=\Delta _{l}-\Delta _{l^{\prime }}$. Therefore, the sign of~$q_{y}$
depends on $\hbar \omega _{c}\left( n-n^{\prime }\right) +\Delta _{2,1}$,
and, if for ordinary impurity scattering there is a current correction
linear in $q_{y}$, it will be an oscillating sign-changing correction. We
shall show that a similar correction to electron conductivity appears when
an additional filling of the second surface subband exceeds that given by
the Boltzmann factor: $N_{2}-N_{1}e^{-\Delta _{2,1}/T_{e}}>0$ . In this
case, the MW affects electron transport in an indirect way, just changing
subband occupation numbers $N_{l}$.

SEs on liquid helium are scattered quasi-elastically by helium
vapor atoms and by capillary wave quanta (ripplons). In this work
we consider only scattering at vapor atoms which is similar to
usual impurity scattering in semiconductor systems. It is
conventionally described by the effective potential $V\left(
\mathbf{R}_{e}-\mathbf{R}_{a}\right) =V_{a}\delta (\mathbf{R}_{e}
- \mathbf{R}_{a})$ and by the interaction Hamiltonian
\begin{equation}
V_{\mathrm{int}}^{(a)}=V_{a}\sum_{\mathbf{K,K}^{\prime }}e^{i\mathbf{KR}
_{e}}a_{\mathbf{K}^{\prime }-\mathbf{K}}^{\dagger }a_{\mathbf{K}^{\prime }},
\label{e1}
\end{equation}%
where $\hbar \mathbf{K}\equiv \{\hbar K_{z},\hbar \mathbf{q}\}$ is
the 3D momentum of a vapor atom, and $a_{\mathbf{K}}^{\dagger }$
($a_{\mathbf{K}}$) is the creation (destruction) operator of vapor
atoms. The 2D vector $\hbar \mathbf{q}$ represents momentum
exchange at a collision.

Considering a multisubband 2D system under the MW resonance
condition, we cannot use the conventional linear-response theory.
As we shall see, even the effective electron temperature
approximation will not help to obtain a sign-changing correction
to $\sigma _{xx}$. For arbitrary population of surface subbands
$n_{l}=N_{l}/N_{e}$, it is convenient to use the force-balance
method (for details, see Ref.~\cite{MonKon-04}). For an isotropic
system, this method assumes that the average kinetic friction $
\mathbf{F}_{\mathrm{scat}}$ acting on electrons is proportional to
the electron current, which can be written as
$\mathbf{F}_{\mathrm{scat} }=-N_{e}m_{e}\nu
_{\mathrm{eff}}\mathbf{v}_{\mathrm{\ av}}$, where $\mathbf{v
}_{\mathrm{av}}$ is the average electron velocity, $N_{e}$ is the
total electron number, $m_{e}$ is the electron mass. The
proportionality factor $ \nu _{\mathrm{eff}}(B)$ represents the
effective collision frequency which generally depends on magnetic
field $B$ (quantum limit) and even on the areal electron density
$n_{s}$ when many-electron effects are considered. Once $ \nu
_{\mathrm{eff}}(B)$ is found, the conductivity tensor can be
obtained from equations similar to the conventional Drude
equations.

For noninteracting electrons, $\mathbf{F}_{\mathrm{scat}}$ can be
found by evaluating the momentum gained by scatterers. Consider an
arbitrary electron distribution $f_{l}\left( \varepsilon \right)
=N_{l}f\left( \varepsilon \right) $, where $\varepsilon $ is the
in-plane energy. We shall use the condition $\int f_{l}\left(
\varepsilon \right) D_{l}\left( \varepsilon \right) d\varepsilon
=N_{l}$, where $D_{l}\left( \varepsilon \right) $ is the
density-of-state function $D_{l}\left( \varepsilon \right)
=-\left(2\pi ^{2}l_{B}^{2}\hbar \right)^{-1}
\sum_{n}\mathrm{Im}G_{l,n}\left( \varepsilon \right) $, and
$G_{l,n}\left( \varepsilon \right) $ is the single-electron
Green's function for the corresponding subband. In the Born
approximation, we obtain
\begin{equation}
\mathbf{F}_{\mathrm{scatt}}=\frac{\hbar ^{2}\nu _{0}^{(a)}}{m_{e}}
N_{e}\sum_{l,l^{\prime }}n_{l}s_{l,l^{\prime
}}\sum_{\mathbf{q}}\mathbf{q} S_{l,l^{\prime }}\left(
\mathbf{q},\omega _{l,l^{\prime }}\right) , \label{e2}
\end{equation}%
where $n_{l}=N_{l}/N_{e}$, $s_{l,l^{\prime }}$ represents $
\sum_{K_{z}}\left\vert \left( e^{iK_{z}z}\right) _{l,l^{\prime
}}\right\vert ^{2}$ normalized according to
Ref.~\cite{MonKonKon-07},
\begin{equation*}
S_{l,l^{\prime }}(\mathbf{q},\omega )=\frac{1}{\pi
^{2}l_{B}^{2}\hbar }\int d\varepsilon f\left( \varepsilon \right)
\sum_{n,n^{\prime }}J_{n,n^{\prime }}^{2}\left( x_{q}\right)
\times
\end{equation*}%
\begin{equation*}
\times \mathrm{Im}G_{l,n}\left( \varepsilon \right)
\mathrm{Im}G_{l^{\prime },n^{\prime }}(\varepsilon +\hbar \omega
),
\end{equation*}%
\begin{equation*}
J_{n,n+m}^{2}\left( x\right) =\frac{n!}{\left( n+m\right)
!}x^{m}e^{-x}\left[ L_{n}^{m}\left( x\right) \right] ^{2},
\end{equation*}%
$L_{n}^{m}\left( x\right) $ is the associated Laguerre
polynomials, $ x_{q}=q^{2}l_{B}^{2}/2$, and $\nu _{0}^{(a)}$ is
the collision frequency in the absence of the magnetic field. If
the collision broadening of Landau levels does not depend on $l$,
the $S_{l,l^{\prime }}(\mathbf{q},\omega )$ coincides with the
dynamical structure factor $S(\mathbf{q},\omega )$ of a
nondegenerate 2D system of noninteracting electrons.

For SEs on liquid helium, electron-electron collision frequency is
much higher than $\nu _{\mathrm{eff}}$. In this case,
$S_{l,l^{\prime }}(\mathbf{q },\omega )$ coincides with its
equilibrium form $S_{l,l^{\prime }}^{(0)}(q,\omega )$ and $f\left(
\varepsilon \right) $ equals to $ Ae^{-\varepsilon /T_{e}}$ only
in the center-of-mass reference frame moving with the drift
velocity $\mathbf{v}_{\mathrm{av}}$ (here $A$ is a normalization
factor). Because of the Galilean invariance along the interface,
in the laboratory frame, $S_{l,l^{\prime }}( \mathbf{q},\omega )$
(as well as the dynamical structure factor~\cite{MonKon-04})
acquires the Doppler shift: $S_{l,l^{\prime }}\left( \mathbf{q}
,\omega \right) =S^{(0)}(q,\omega -\mathbf{q\cdot
v}_{\mathrm{av}})$. The $\hbar \mathbf{q\cdot v}_{\mathrm{av}}$
represents an additional energy exchange at a collision which
appears in the center-of-mass reference frame.

In the cumulant approach~\cite{Ger-76}, $\mathrm{Im}G_{l,n}\left(
\varepsilon \right) $ has a Gaussian shape with the Landau level broadening $%
\Gamma _{l}$ independent of $n$ (for the lowest surface subband, $\Gamma
_{1}= \sqrt{\frac{2}{\pi }\hbar ^{2}\omega _{c}\nu _{0}^{(a)}}$). In this
case, \
\begin{equation}
S_{l,l^{\prime }}^{(0)}\left( q,\omega \right) =\frac{2\pi ^{1/2}\hbar }{
Z_{\Vert }}\sum_{n,n^{\prime }}\frac{J_{n,n^{\prime }}^{2}\left(
x_{q}\right) }{\Gamma _{l,l^{\prime }}}e^{-n\hbar \omega
_{c}/T_{e}}I_{l,l^{\prime };n^{\prime }-n}\left( \omega \right) ,  \label{e3}
\end{equation}%
where $Z_{\Vert }^{-1}=1-e^{-\hbar \omega _{c}/T_{e}}$,
\begin{equation*}
I_{l,l^{\prime };m}\left( \omega \right) =\exp \left[ -\left(
\frac{\hbar \omega -m\hbar \omega _{c}-\Gamma
_{l}^{2}/4T_{e}}{\Gamma _{l,l^{\prime }}} \right)
^{2}+\frac{\Gamma _{l}^{2}}{8T_{e}^{2}}\right] ,
\end{equation*}%
and $\Gamma _{l,l^{\prime }}^{2}=\left( \Gamma _{l}^{2}+\Gamma
_{l^{\prime }}^{2}\right) /2$. We assume that $\Gamma _{l}\ll T$.

Using $S_{l,l^{\prime }}^{(0)}\left( q,-\omega \right) =e^{-\hbar \omega
/T_{e}}S_{l^{\prime },l}^{(0)}\left( q,\omega \right) $, Eq.~(\ref{e2}) \
can be represented as
\begin{equation*}
\mathbf{F}_{\mathrm{scatt}}=\frac{\hbar ^{2}\nu _{0}^{(a)}}{2m_{e}}%
N_{e}\sum_{l,l^{\prime }}s_{l,l^{\prime }}\sum_{\mathbf{q}}\mathbf{q}%
S_{l,l^{\prime }}^{(0)}\left( q,\omega _{l,l^{\prime }}-\mathbf{q\cdot v}_{%
\mathrm{av}}\right) \times
\end{equation*}%
\begin{equation}
\times \left[ n_{l}-n_{l^{\prime }}e^{-\Delta _{l,l^{\prime
}}/T_{e}}e^{\hbar \mathbf{q}\cdot \mathbf{v}_{\mathrm{av}}/T_{e}}\right] .
\label{e4}
\end{equation}%
For slow drift velocities, $\hbar \mathbf{q\cdot
v}_{\mathrm{av}}\ll \Gamma _{l,l^{\prime }}$ and $\hbar
\mathbf{q\cdot v}_{\mathrm{av}}\ll T_{e}$, linear in
$\mathbf{v}_{\mathrm{av}}$ terms of $\mathbf{F}_{\mathrm{scatt}}$
and the effective collision frequency $\nu _{\mathrm{eff}}$ can be
easily obtained from Eq.~(\ref{e4}). The linear term obtained
expanding $e^{\hbar \mathbf{q\cdot v}_{\mathrm{av}}/T_{e}}$
represents the usual result of which the effective collision
frequency is always positive.

There is another term originated from the expansion of $\left(
n_{l}-n_{l^{\prime }}e^{-\Delta _{l,l^{\prime }}/T_{e}}\right) \mathbf{q}%
S_{l,l^{\prime }}^{(0)}\left( q,\omega _{l,l^{\prime }}-\mathbf{q\cdot v}_{%
\mathrm{av}}\right) $ in $\hbar \mathbf{q\cdot v}_{\mathrm{av}}/\Gamma
_{l,l^{\prime }}$. This is the linear in $\mathbf{q}$ term we were searching
for. It is similar to the corresponding term of Ref.~\cite{Ryz-69} obtained
for photon-induced impurity scattering because $S_{l,l^{\prime
}}^{(0)}\left( q,\omega _{l,l^{\prime }}-\mathbf{q\cdot v}_{\mathrm{av}%
}\right) $ initially contained the $\delta $-function representing energy
conservation. Expanding $I_{l,l^{\prime };m}\left( \omega _{l,l^{\prime }}-%
\mathbf{q\cdot v}_{\mathrm{av}}\right) $ of Eq.~(\ref{e4}), we obtain the
factor $\Delta _{l,l^{\prime }}-\left( n^{\prime }-n\right) \hbar \omega
_{c}-\Gamma _{l}^{2}/4T_{e}$ which has different signs at the opposite sides
of the point where it equals zero.

If heating of the electron system is small ($T_{e}\ll \Delta _{2,1}$), we
can use the two-subband model ($l=1,2$) and find the sign-changing
correction to $\nu _{\mathrm{eff}}$
\begin{equation*}
\delta \nu _{\mathrm{eff}}=-\frac{\nu _{0}^{(a)}\hbar ^{2}\omega _{c}^{2}}{%
\sqrt{\pi }}\frac{s_{1,2}}{\Gamma _{2,1}^{2}}\left( n_{2}-n_{1}e^{-\Delta
_{2,1}/T_{e}}\right) \times
\end{equation*}%
\begin{equation}
\times \left[ \coth \left( \frac{\hbar \omega _{c}}{2T_{e}}\right) \Phi
\left( \omega _{c}\right) +\Theta \left( \omega _{c}\right) \right] ,
\label{e5}
\end{equation}%
where
\begin{equation*}
\Phi \left( \omega _{c}\right) =\delta F_{2,1}\left( \omega _{c}\right)
-e^{\Delta _{2,1}/T_{e}}\delta F_{1,2}\left( \omega _{c}\right) \text{ \ \ }
\end{equation*}%
\begin{equation*}
\Theta \left( \omega _{c}\right) =\delta H_{2,1}\left( \omega _{c}\right)
-e^{\Delta _{2,1}/T_{e}}\delta H_{1,2}\left( \omega _{c}\right) .
\end{equation*}%
For $\Delta _{l,l^{\prime }}>0$,
\begin{equation*}
\delta F_{l,l^{\prime }}\left( \omega _{c}\right) =\sum_{m=0}^{\infty
}I_{l,l^{\prime };m}\left( \Delta _{l,l^{\prime }}\right) \frac{\left(
\Delta _{l,l^{\prime }}-m\hbar \omega _{c}-\Gamma _{l}^{2}/4T_{e}\right) }{%
\Gamma _{l,l^{\prime }}}.
\end{equation*}%
In the opposite case ($\Delta _{l,l^{\prime }}<0$),%
\begin{equation*}
\delta F_{l,l^{\prime }}\left( \omega _{c}\right) =\sum_{m=0}^{\infty
}e^{-m\hbar \omega _{c}/T_{e}}I_{l,l^{\prime };-m}\left( \Delta
_{l,l^{\prime }}\right) \times
\end{equation*}%
\begin{equation*}
\times \frac{\left( \Delta _{l,l^{\prime }}+m\hbar \omega
_{c}-\Gamma _{l}^{2}/4T_{e}\right) }{\Gamma _{l,l^{\prime }}}.
\end{equation*}%
The corresponding form of $\delta H_{l,l^{\prime }}\left( \omega
_{c}\right) $ differs from that of $\delta F_{l,l^{\prime }}\left(
\omega _{c}\right) $ only by the additional factor $m$ inside the
sum. One can see that $\delta F_{2,1}$
and $\delta F_{1,2}$ entering $\Phi \left( \omega _{c}\right) $ [as well as $%
\delta H_{2,1}$ and $\delta H_{1,2}$ \ in $\Theta \left( \omega _{c}\right) $%
] have opposite signs. This means that the contribution of scattering from $%
l=2$ to $l=1$ is not cancelled by the contribution of scattering from $l=1$
to $l=2$.

The conventional contribution to $\nu _{\mathrm{eff}}$ originated from the
expansion of $e^{\hbar \mathbf{q\cdot v}_{\mathrm{av}}/T_{e}}$ can be
written as%
\begin{equation*}
\nu \left( B\right) =\frac{\nu _{0}^{(a)}\hbar ^{2}\omega _{c}^{2}}{2\pi
^{1/2}T_{e}}\sum_{l,l^{\prime }}\frac{s_{l,l^{\prime }}}{\Gamma
_{l,l^{\prime }}}n_{l^{\prime }}e^{-\Delta _{l,l^{\prime }}/T_{e}}\times
\end{equation*}%
\begin{equation}
\times \left[ \coth \left( \frac{\hbar \omega _{c}}{2T_{e}}\right)
F_{l,l^{\prime }}\left( \omega _{c}\right) +H_{l,l^{\prime
}}\left( \omega _{c}\right) \right] ,   \label{e6}
\end{equation}%
where $F_{l,l^{\prime }}$ and $H_{l,l^{\prime }}$ are defined
similar to $ \delta F_{l,l^{\prime }}$ and $\delta H_{l,l^{\prime
}}$ with the exception that their equations do not contain the
factor $\left( \Delta _{l,l^{\prime }}\mp m\hbar \omega
_{c}-\Gamma _{l}^{2}/4T_{e}\right) /\Gamma _{l,l^{\prime }}$ . For
SEs on liquid helium, Landau levels are extremely narrow: $\Gamma
_{l}\ll T$. In this case, the collision frequency term of
Eq.~(\ref{e5}) originated from the expansion in $\hbar
\mathbf{q\cdot v}_{\mathrm{av} }/\Gamma _{l,l^{\prime }}$ acquires
an additional large factor $T_{e}/\Gamma _{l,l^{\prime }}\gg 1$ as
compared to the conventional term of Eq.~(\ref{e6}). Therefore,
even a small nonequilibrium filling of the excited subband $
n_{2}-n_{1}e^{-\Delta _{2,1}/T_{e}}>0$ can lead to giant
oscillations in $ \nu _{\mathrm{eff}}$.

\begin{figure}[tbp]
\begin{center}
\includegraphics[width=9.5cm]{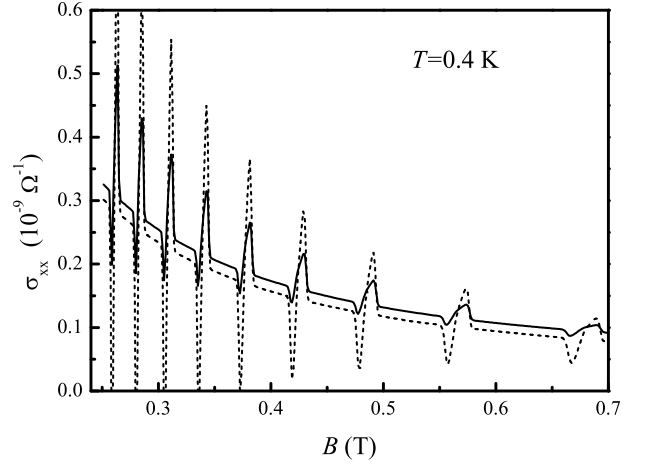}
\end{center}
\caption{Magnetoconductivity vs magnetic field for two MW power levels: $%
\Omega _{R}/ \nu _{0}^{(a)}=0.2$ (solid) and $\Omega _{R}/ \nu
_{0}^{(a)}=0.4$ (dashed). } \label{f1}
\end{figure}

The results of numerical evaluations of $\sigma _{xx}$, taking
into account both the usual contribution to $\nu _{\mathrm{eff}}$
defined by Eq.~(\ref{e6}) and the sign-changing correction of
Eq.~(\ref{e5}), are shown in Fig.~\ref{f1} for two levels of MW
power defined by the ratio of the Rabi frequency $\Omega _{R}$ to
$\nu _{0}^{(a)}$. Fractional occupancies $n_{l}$ are found from
the rate equation of the two-subband model, while electron
temperature $T_{e}$ is obtained from the energy balance equation.
The MW resonance frequency equals $95\,\mathrm{GHz}$. Other
parameters correspond to experiments with liquid $^{3}\mathrm{He}$
at $T=0.4\,\mathrm{K}$. One can see that for noninteracting
electrons, the amplitude of the sign-changing correction to
$\sigma _{xx}$ becomes larger in the low magnetic field range,
and, at high MW powers, $\sigma _{xx}$ becomes negative. The shape
of $ \sigma _{xx}$ oscillations obtained here is in striking
accordance with experimental data observed for SEs on liquid
helium~\cite{KonKon-2010}. Conductivity minima occur when the
parameter $\Delta _{2,1}/\hbar \omega _{c} $ is slightly above an
integer, which also agrees with the observation.

\begin{figure}[tbp]
\begin{center}
\includegraphics[width=9.5cm]{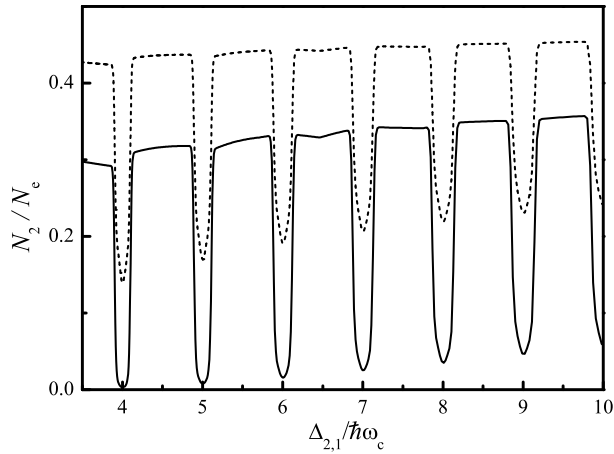}
\end{center}
\caption{The fractional occupancy of the second subband for two MW
power levels: $\Omega _{R}/ \nu _{0}^{(a)}=0.2$ (solid) and
$\Omega _{R}/ \nu _{0}^{(a)}=0.4$ (dashed). } \label{f2}
\end{figure}

The fractional occupancy of the second subband $n_{2}$ as a
function of the parameter $\Delta _{2,1}/\hbar \omega _{c}$ is
shown in Fig.\ref{f2}. It is reduced in the vicinity of the
condition $\Delta _{2,1}/\hbar \omega _{c}\rightarrow m$ (here $m$
is an integer) due to elastic inter-subband scattering.

The many-electron effect can be taken into account considering an electron
moving in a quasi-uniform electric field of other electrons $E_{f}$ due to
thermal fluctuations~\cite{DykKha-79}. In this case, according to Ref.~\cite%
{MonKon-04}, $\Gamma _{l,l^{\prime }}$ entering $S_{l,l^{\prime
}}^{(0)}\left( q,\omega \right) $ should be replaced by
$\sqrt{\Gamma _{l,l^{\prime }}^{2}+x_{q}\Gamma _{C}^{2}},$ where
$\Gamma _{C}=\sqrt{2}
eE_{f}^{(0)}l_{B}$, and $E_{f}^{(0)}\simeq 3\sqrt{T_{e}}n_{s}^{3/4}$. Since $%
l_{B}\propto 1/\sqrt{B}$, the many-electron effect broadens
magnetooscillations and reduces their amplitudes in the low field range,
which is also in accordance with experimental observations.

It should be noted that $\sigma _{xx}(B)$ curves do not change
much even if we set $T_{e}=T$. Nevertheless, electron temperature
as a function of the parameter $\Delta _{2,1}/\hbar \omega _{c}$
oscillates and rises when $ \Delta _{2,1}/\hbar \omega
_{c}\rightarrow m $. In this case, quasi-elastic decay transfers
$\Delta _{2,1}$ to the kinetic energy of the in-plane motion. At
$\Omega _{R}/\nu _{0}^{(a)}\sim 0.2$, the maximum increase in $
T_{e}$ is less than $2\,\mathrm{K}$ which allows to use the
two-subband approximation. For high MW powers, under the condition
$\Delta _{2,1}/\hbar \omega _{c}\rightarrow m$ electron population
of higher subbands should be taken into account. Electron heating
can be suppressed by increasing the holding electric field
$E_{\bot }$ or by lowering the ambient temperature $T$. In the
later case, electrons are predominately scattered by ripplons.
Nevertheless, qualitatively magnetooscillations of $\sigma _{xx}$
should be the same because typical energy of ripplons involved is
extremely small. There is also a possibility to use a higher MW
resonance condition $\hbar \omega =\Delta _{l}-\Delta _{1}$ with
$l>2$.

Concluding, we have shown that remarkable MW-induced oscillations
of magnetoconductivity of SEs on liquid helium and zero-resistance
states observed in Ref.~\cite{KonKon-2009,KonKon-2010} can be
understood as a consequence of inter-subband electron scattering
by vapor atoms or ripplons under the condition that the fractional
occupancy of the excited subband exceeds that given by the
conventional Boltzmann factor. This explains why the MW resonance
condition $\hbar \omega =\Delta _{2}-\Delta _{1}$\ is crucial for
observation of magnetooscillations and ZRS in this system.
Electron-electron interaction, increasing with electron density,
is shown to suppress the amplitude of magnetooscillations and to
destroy the ZRS.

The author is grateful to K. Kono and D. Konstantinov for acquainting with
experimental data before publication and for helpful discussions.

\end{document}